\documentclass[aps,prl,twocolumn,groupedaddress,amsmath,amssymb,showpacs]{revtex4}
\usepackage[dvips]{graphicx}
\begin{document}

\title{Cold Collision Frequency Shift in Two-Dimensional Atomic Hydrogen}

\author{J. Ahokas, J. J{\"a}rvinen,  and S. Vasiliev}
\email{servas@utu.fi} \affiliation { Wihuri Physical Laboratory,
Department of Physics, University of Turku, 20014 Turku,
Finland\\}

\date{\today}

\begin{abstract}
We report a measurement of the cold collision frequency shift in
atomic hydrogen gas adsorbed on the surface of superfluid $^4$He
at $T\lesssim 90$ mK. Using two-photon electron and nuclear
magnetic resonance in 4.6 T field we separate the resonance line
shifts due to the dipolar and exchange interactions, both
proportional to surface density $\sigma$. We find the clock shift
$\Delta \nu_{c}=-1.0(1)\times 10^{-7}$ Hz$\cdot$
cm$^{-2}\cdot\sigma$, which is about 100 times smaller than the
value predicted by the mean field theory and known scattering
lengths in the 3D case.
\end{abstract}

\maketitle

Elastic collisions between atoms play an important role in the
physics of quantum gases. At low enough temperatures, when the
thermal de Broglie wavelength $\Lambda_{th}=(2\pi\hbar^2/ m k_B
T)^{-1/2}$ exceeds the range of the interaction potential, one
enters the cold collision regime. In this case the effect of
collisions can be easily interpreted and leads to a shift of the
resonance lines known as cold collision or clock shift. This
phenomenon puts severe limitations on the precision of atomic
clocks \cite{Gibble}. The cold collision frequency shift of the
1$S$-2$S$ transition in three-dimensional (3D) atomic hydrogen gas
\cite{Killian98} has been used to detect the formation of the
Bose-Einstein condensate \cite{Fried98}. Recently the studies of
collisions in quantum degenerate gases of reduced dimensionality
\cite{Grimm04, Dalibard06} attracted considerable interest and
novel phenomena like modified interaction properties have been
predicted \cite{Petrov04}.

The interaction potentials between two hydrogen atoms are well
described by theory and allow accurate calculations of collisional
cross sections \cite{Stoof88, Kokkelmans97}. However, serious
discrepancies between the theory and experimental studies of
collisions of hydrogen atoms in the ground electronic state remain
up to date \cite{Hayden96, Kokkelmans97}. Mainly due to their
application to a hydrogen maser, the experiments were performed in
nearly zero field. In this case the frequency shifts depend in a
complicated way on the occupations of the hyperfine states of the
colliding atoms which are nearly equal in thermal equilibrium even
at the lowest temperatures of 0.5 K. This complication will be
eliminated in a strong magnetic field when one can easily realize
a situation of electron and nuclear polarized gas
(H\makebox[1.01ex][l]{$\downdownarrows$}-), with $>99\%$ of atoms
being in a single hyperfine state \cite{BlueBible}. In this case
broadening of the resonance lines due to field inhomogeneities
makes precise spectroscopic measurements difficult and it requires
much higher densities of the gas to discern collisional shifts.
The problem can be solved in atomic hydrogen gas adsorbed on the
surface of helium film. The surface binding potential effectively
compresses atoms, strongly increasing the collision rate and the
interaction energy.

Relatively large (up to 5 G) shifts of the electron spin resonance
(ESR) lines of the adsorbed atoms from that of the 3D gas have
been observed \cite{Shinkoda91, Vasilyev02}. The shifts are
proportional to the surface density $\sigma$ and anisotropic with
respect to the orientation of the surface in the polarizing
magnetic field, in fair agreement with calculations of the
internal dipolar fields in the 2D gas \cite{Shinkoda91}. A
possibility of the clock shift was not considered as it was
believed that the shift may not appear due to a coherent
interaction of the atoms with the rf excitation so that the
interaction in collisions always occurs via triplet potential.
Later the influence of coherence has been clarified in experiments
with cold bosonic and fermionic alkali vapors \cite{Harber02,
Zwierlein03}. It has been proved \cite{Zwierlein03} that the
interaction of rf excitation with atomic system is always coherent
and the clock shift does not depend on the level of coherence
between the internal states.

Here we present the first measurement of the exchange and dipolar
contributions to the interaction energy in
H\makebox[1.01ex][l]{$\downdownarrows$}- gas adsorbed on the
surface of superfluid helium. In addition to the ESR method used
previously \cite{Jarvinen05}, we employed nuclear magnetic
resonance (NMR), for which there is no clock shift because atoms
in the initial and final states interact via the same triplet
potential. We carried out a coherent two-photon study of the
resonances as well as a separate measurement of ESR and NMR
transitions in the sample of the same surface density. Results of
the measurements agree well with each other.

There is a single bound state for
H\makebox[1.01ex][l]{$\downdownarrows$}- on helium surface with
the adsorption energy of $E_a=1.14(1)$ K. Due to the low mass $m$
and low $E_a$ delocalization of H$\downarrow$ in the out-of-plane
direction $l=\hbar/\sqrt{2E_am}\approx0.5$ nm largely exceeds the
three-dimensional $s$-wave scattering length $a_t\approx72$ pm.
Therefore, in terms of collisions the gas may be considered as
three-dimensional. In the limit $l\gg a_t$ the interaction energy
per particle of such a quasi 2D gas can be evaluated using a
scaling approach with the effective bulk density
$n\approx\sigma/l$ \cite{Kagan82,Petrov04}:
\begin{equation}
E_i \approx \frac {4\pi \hbar^2 g a_t \sigma} {m l}, \label{Ei}
\end{equation}
where $g$ is the two body correlator equal to 2 in the thermal,
non-condensed gas of bosons \cite{Harber02, Zwierlein03}.
Densities exceeding $\sigma=10^{12}$ cm$^{-2}$ are easily
accessible by the thermal compression method \cite{Jarvinen05}.
Then, the mean field interaction energy is of the same order as in
the bulk gas of density $n\sim\sigma/l \approx2\times10^{19}$
cm$^{-3}$, ten orders of magnitude larger than in a typical
hydrogen maser.

As described in ref. \cite{Jarvinen05}, two-dimensional samples of
atomic hydrogen gas are created on a cold spot (CS) of the sample
cell (SC) (fig. 1 \textit{a}). In this work we keep the hydrogen
dissociator continuously filling the SC. To reduce the flux of
atoms in the highly reactive state \textit{a} (fig. 1 \textit{b})
we feed the atoms into the SC through a beam polarizer. It is a
separate chamber with a surface area of $\sim10$ cm$^2$, located
in a reduced magnetic field ($\approx3.0$ T) at a temperature of
100...130 mK, optimized for preferential recombination of the
\textit{a} state. Varying the temperatures of the SC and CS and
the dissociator power we get a steady state with bulk
H\makebox[1.01ex][l]{$\downdownarrows$}- density
$n\lesssim10^{13}$ cm$^{-3}$ in the SC and surface density
$\sigma=10^{11}-3.5\times 10^{12}$ cm$^{-2}$ on the CS. The main
advantage of this method is a long term stability of the surface
density and temperature, required for accurate spectroscopic
studies. In addition to the 128 GHz Fabry-Perot ESR resonator, we
have an NMR resonator in the vicinity of the CS (fig. 1
(\textit{a})). It is a helix (HR) resonant at 909 MHz, close to
the \textit{a-b} nuclear spin transition (fig. 1 (\textit{b})) in
the field of $B_0\approx4.6$ T.
\begin{figure}
\includegraphics[width=8 cm]{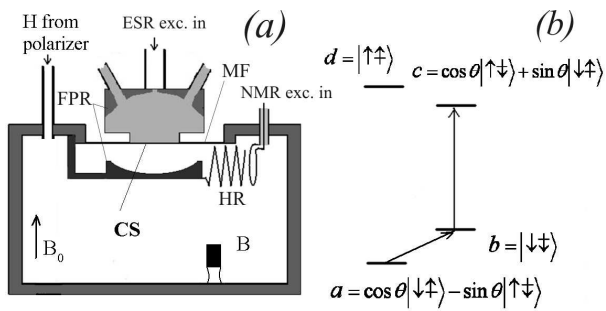}
\caption{\textit{(a)} - Schematic drawing of the sample cell. FPR
= Fabry-Perot ESR resonator, HR = helical NMR resonator, MF =
Mylar foil, B = bolometer, CS = cold spot. \textit{(b)} -
hyperfine level diagram for hydrogen atom in a strong magnetic
field. Arrows denote electron and nuclear spin projections.}
\label{cell}
\end{figure}

We performed continuous wave ESR on bulk and surface
H\makebox[1.01ex][l]{$\downdownarrows$}- sweeping the magnetic
field across the \textit{b-c} transition at a fixed frequency
$\nu_{bc}$. Both absorption and dispersion components of the
signal are detected simultaneously by a cryogenic millimeter-wave
receiver \cite{OurESR}. The ESR excitation power is always kept
low enough to avoid distortions of the resonance lines
\cite{Vasilyev02}. A typical \textit{b-c} spectrum shown in (fig.
2 (a)) contains two peaks.
\begin{figure}
\includegraphics[width=8 cm]{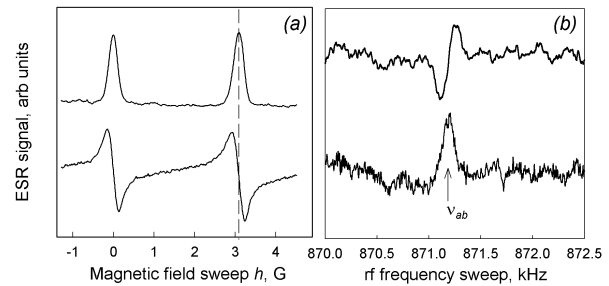}
\caption{Observed ESR and two-photon spectra. Upper traces -
absorption, lower traces - dispersion. \textit{(a)} - bulk gas
(left peak) and surface (right peak) ESR lines; \textit{(b)} -
two-photon resonances.} \label{spectra}
\end{figure}
One originates from free atoms in the bulk gas, and another - from
adsorbed atoms. The distance between the peaks has been found to
be proportional to the surface density \cite{Shinkoda91,
Vasilyev04}. The much weaker \textit{a-b} transition was studied
calorimetrically with a miniature bolometer utilizing the large
recombination energy released after the transfer of
H\makebox[1.01ex][l]{$\downdownarrows$}- atoms to the reactive
\textit{a}-state \cite{Safonov98}.

We employed the double electron-nuclear resonance method to study
adsorbed H\makebox[1.01ex][l]{$\downdownarrows$}- gas. The sweep
field $h$ of the ESR spectrometer was stopped in the center of the
\textit{b-c} transition for the surface atoms (dashed line in fig.
2 (a)) and then rf excitation was applied to the NMR helix. By
sweeping the rf frequency we detect the \textit{a-b} transition
through a change in the ESR signal (fig. 2 (b)), which can be
caused by two effects: (i) Destruction of the \textit{b}-state
population on the CS due to the transfer of atoms to the state
\textit{a} with subsequent recombination. This is similar to the
classical electron-nuclear double resonance (ENDOR) well known in
the magnetic resonance spectroscopy of solids. (ii) Coherent
two-photon excitation of three level system. In (i) the effect of
the \textit{b} state destruction is strongly enhanced by the
change of the surface line position which leads to a decrease in
the absorption and dispersion components of the ESR signal. The
$b\rightarrow a$ transition is also detected by the bolometer
registering the concomitant recombination heat. The $\approx600$
Hz width of the ENDOR line is defined by the magnetic field
inhomogeneity over the CS. We observed such ENDOR spectra at high
enough ($\gtrsim100$ mK) temperatures, when a rapid exchange with
the bulk gas destroys the coherence of the adsorbed atoms with the
rf excitation. At lower temperatures the surface residence time
becomes long enough to maintain coherence and the spectra look
completely different (fig. 2 (b)). The sign of the absorption
component change depends on the detuning and the signal has a
"dispersive" lineshape, while dispersive component looks like
"absorption". The 120 Hz line width is much smaller than that of
the ENDOR line. We verified by changing the rf power that the two
photon line shapes and positions are not influenced by the
nonlinear effects caused by too strong excitation. The bolometer
does not detect any extra recombination heat, implying that all
atoms remain in the \textit{b} state. Similar effects are known in
the field of quantum optics when a two-photon field interacts
coherently with a three level system \cite{QOBook}. They are
beyond the scope of the present work and require further studies.
We used the two-photon resonance to measure the \textit{a-b}
transition frequency as a function of surface density on the CS.

There are two reasons for the density dependent shift of the
resonance lines. First is the average dipolar field $B_d$ in the
plane of the adsorbed atoms, which has negative sign for the
surface perpendicular to the polarizing field. Second is the clock
shift, following from eq. \ref{Ei}
\begin{equation}
\Delta \nu_c= \frac {2 \hbar g} {m l}
(a_s-a_t)\cdot\sigma\equiv\kappa\cdot\sigma, \label{dnuc}
\end{equation}
where $a_t$ and $a_s$ are the triplet and singlet s-wave
scattering lengths. In addition, there is also a density
independent frequency shift due to the change $\Delta A_w$ of the
hyperfine constant $A$ caused by the interaction of the adsorbed
atoms with the substrate, known as the wall shift (WS)
\cite{Zitzewitz71}. Taking the above mentioned shifts into
account, the resonance condition for the \textit{b-c} transition
of atoms adsorbed on the surface can be written in the strong
field approximation as
\begin{equation}
\nu_{bc}= \frac {\gamma_e}{2\pi}(B_0+h+B_d)- \frac {A +\Delta
A_w}{2}+ \Delta \nu_c, \label{ESR}
\end{equation}
where $h$ is the sweep field and $B_0$ is the main polarizing
field.  We adjust the latter so that the resonance line of the
bulk gas is observed at $h=0$, i.e. $\nu_{bc}= \gamma_e B_0
/2\pi-A/2$. According to eq. \ref{ESR} the position of the surface
ESR line $h_e=\alpha\cdot\sigma$ depends on the internal dipolar
field $B_d$ and the clock shift $\Delta \nu_c$, both proportional
to $\sigma$. The clock shift for the \textit{a-b} transition is
absent and the resonance equation is
\begin{equation}
\nu_{ab}= \frac {\gamma_n}{2\pi}(B_0+h+B_d)+ \frac {A +\Delta
A_w}{2}. \label{NMR}
\end{equation}
In the two-photon experiment we measure $\nu_{ab}$ with the sweep
field being fixed to the surface resonance, i.e. $h=h_e$. Then,
denoting the resonance frequency of the \textit{a-b} transition
for the bulk gas as $\nu_{ab}^0 \equiv\gamma_n B_0/2 \pi+ A/2$ and
neglecting $\gamma_n /\gamma_e$ with respect to unity,  we obtain
from eqs. \ref{dnuc}-\ref{NMR}
\begin{equation}
\nu_{ab}-\nu_{ab}^0= \frac {\Delta A_w}{2}-
\frac{\gamma_n}{\gamma_e}\Delta\nu_c= \frac {\Delta
A_w}{2}-\frac{\gamma_n}{\gamma_e} \frac{\kappa}{\alpha}h_e.
\label{CS}
\end{equation}

In fig. 3 we plot the frequency difference $\nu_{ab}-\nu_{ab}^0$
as a function of the ESR line shift $h_e$. Fitting the data to a
straight line we obtain the wall shift $\Delta A_w=-45.58(4)$ kHz
and the clock shift $\Delta \nu_c=-1.0(1)\times 10^{-6}$ Hz cm$^2
\cdot \sigma$, where we used known relation $h_e= \alpha \cdot
\sigma$ with $\alpha=1.1(1)\times 10^{-12}$ G$\cdot$cm$^2$
\cite{Vasilyev04}. The latter is the main reason for the 10\%
systematic error of the clock shift which is much larger than the
other errors marked with bars in fig. 3. The contribution of the
clock shift is $\Delta B_c=2\pi \Delta \nu_c
/\gamma_e=3.6(4)\times 10^{-13}$ G cm$^2 \cdot \sigma$, which
comprises approximately one third of the total observed ESR line
shift $\Delta B$, the rest being due to the internal dipolar field
$B_d=7.4(7)\times 10^{-13}$ G cm$^2 \cdot \sigma$.
\begin{figure}
\includegraphics[width=8 cm]{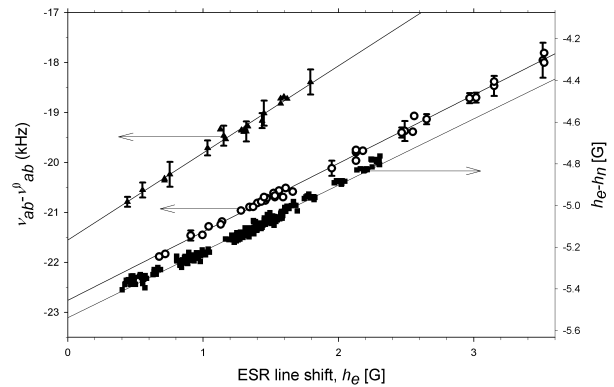}
\caption{Frequency shift of the \textit{a-b} transition as a
function of the ESR line shift obtained by the two-photon method:
$\circ$ - pure $^4$He film, $\blacktriangle$ - $^3$He-$^4$He
mixture. $\blacksquare$ - difference of the ESR and NMR resonance
line positions measured at fixed frequencies $\nu_{bc}$ and
$\nu_{ab}$. Solid lines are linear fits to the data. } \label{CS}
\end{figure}

The clock shift can be measured by a somewhat different method. We
apply both ESR and NMR excitations simultaneously, with the latter
tuned to the resonance value of the free atoms in the field $B_0$,
i.e. we set $\nu_{ab}=\nu_{ab}^0$. Sweeping magnetic field we
detect both \textit{b-c} and \textit{a-b} transitions by the
bolometer and study their positions $h_e$ and $h_n$ as functions
of the surface density, or the ESR line shift $h_e$. If the
density is uniform over the CS, the difference of the surface line
positions is proportional to the clock shift. For these
experimental conditions we get from eqs. 3 and 4
\begin{equation}
\frac{\gamma_n}{2\pi}(h_n-h_e)= \frac {\Delta A_w}{2} - \frac
{\gamma_n}{\gamma_e}\Delta \nu_{c}= \frac {\Delta
A_w}{2}-\frac{\gamma_n}{\gamma_e} \frac{\kappa}{\alpha}h_e.
\label{hn}
\end{equation}

Data for such an experiment are also plotted in fig. 3. The points
follow straight line with the same (within $\lesssim3\%$) slope as
the two-photon data and shifted down by $\approx0.1$ G, which can
be explained by the magnetic field difference between the edge and
the center of the CS. The ESR transition is effectively excited in
the $\sim1$ mm diam. field maximum in the CS center. The rf field
is larger near the CS edge, which is closer to the HR. Therefore,
the NMR and ESR probe the adsorbed gas in slightly different
positions. This makes a major difference with the two-photon
measurement where ESR and NMR photons are absorbed
\textit{simultaneously} by the atoms located in the CS center.
Good agreement with the two photon data can be considered as a
confirmation of the uniform density over the cold spot and of the
reliability of the clock shift measurement.

Although we cannot directly measure the surface gas temperature,
its upper limit of 70...90 mK can be estimated from the adsorption
isotherm assuming dynamic equilibrium between the bulk and surface
gases \cite{Vasilyev04}. The lower bound is set by the temperature
of the $^3$He-$^4$He mixture cooling the CS, which was varied in
the range 50...90 mK. We have not observed any influence of the
surface temperature on $\nu_{ab}$ in this temperature range.

We compare our data for the wall shift with the result $\Delta
A_w=-44.8(10)$ kHz of ref. \cite{Pollack92} measured for the
surface parallel to the magnetic field, -43.2(1) kHz
 for the surface having no preferential orientation \cite{Pollack86} and $\Delta
A_w=-49$ kHz in zero field \cite{Morrow81}. The agreement is good,
especially taking into account the very different methods used in
these measurements.

Using our data for the clock shift and eq. \ref{CS} we obtain the
difference of the scattering lengths $a_t-a_s=0.42(4)$ pm. The
theoretical values of $a_t=72$ pm and $a_s=17$ pm
\cite{Williams93} give a much larger difference $a_t-a_s=55$ pm.
Our result may be interpreted as the singlet scattering length
being nearly equal to the triplet one, i.e. 4 times larger than
predicted. The authors of ref. \cite{Williams93} pointed out that
their calculation for $a_s$ is sensitive to nonadiabatic
corrections. But the $\sim$5 $\%$ difference they obtained
completely neglecting these corrections can not explain the value
of $a_s$ extracted from our data.

The very small value of the clock shift found in this work can be
due to the change of the collision properties in the adsorbed
phase. To verify such possibility we attempted to modify the
interaction of the adsorbed atoms with the film by admixing some
small amount of $^3$He. The presence of $^3$He on the surface
reduces the adsorption energy and increases delocalization $l$ of
the adsorbed atoms \cite{Safonov01}. This should lead to a $\sim
l^{-3}$ reduction of the wall shift \cite{Zitzewitz71} and to a
$\sim l^{-1}$ reduction of the clock shift (eq. \ref{dnuc}). We
condensed $1.5(5)\times10^{16}$ of $^3$He atoms into the SC, which
would provide a maximum surface coverage of $1.5\cdot10^{14}$
cm$^{-2}$, assuming that all the $^3$He atoms are evenly
distributed on the surface. With this coverage the adsorption
energy decreases to $\approx0.9$ K \cite{Safonov01} and
corresponding increase of $l$ is $\lesssim10\%$. The actual
concentration of $^3$He on the CS is difficult to estimate since
it strongly depends on the temperature profile inside the SC.
However, we believe that the concentration did not change in the
experiment, since we observed no difference of $\nu_{ab}$ at
different SC temperatures (70-90 mK) and same $h_e$. In the
measurements (fig. 3) we found a $\approx6\%$ \textit{decrease} of
the wall shift implying a larger $l$, i.e. in qualitative
agreement with the expectation. The change of the data slope, on
the contrary to the prediction of eq. 1, corresponds to a $25\%$
\textit{increase} of the clock shift. Unfortunately, we were not
able to increase $l$ further. Adding more $^3$He created
instabilities of the helium level in the SC and no reliable data
could be taken.

The observed increase of the clock shift at larger $l$ is in line
with our previous measurements of the surface three-body
recombination rate constant \cite{Vasilyev04}. The rate constant
found in that work has also revealed serious discrepancy with the
estimate based on the scaling approach.

In conclusion, we have measured the clock shift of atomic hydrogen
gas adsorbed on the surface of helium film. The observed value is
$\sim$100 times smaller than evaluated on the basis of the mean
field theory applied to a quasi 2D gas and existing data on the
scattering lengths in the 3D case. A small increase of the
delocalization length of adsorbed atoms leads to a substantial
increase in the clock shift. This may indicate that the strong
reduction of the clock shift is a feature of the reduced
dimensionality. A measurement of the clock shift in the 3D case at
high magnetic field would help to understand the nature of the
discrepancy. The two-photon magnetic resonance method realized in
this work opens up new possibilities for studying the coherent
interaction of the rf excitation with the quantum degenerate gas
of H.

\begin{acknowledgments}
This work was supported by the Academy of Finland grant 206109 and
the Wihuri Foundation. We also wish to thank K.-A. Suominen, S.
Jaakkola and M. Hayden for valuable discussions.
\end{acknowledgments}

\bibliography{ClockShift}

\end{document}